\documentclass[prb,twocolumn,epsfig,superscriptaddress]{revtex4}
\usepackage{graphicx}
\usepackage{psfrag}

\begin{document} 
\title{
Effective screening of localized charged perturbations \\
in metallic nanotubes: roles of massive bands
}
\date{\today}

\author{K. Sasaki}
\affiliation{Department of Physics, Tohoku University, Sendai 980-8578,
Japan} 
\author{A. A. Farajian}
\author{H. Mizuseki}
\author{Y. Kawazoe}
\affiliation{Institute for Materials Research, Tohoku University, Sendai
980-8577, Japan}

\begin{abstract}
 The massive-band effects on screening behavior of metallic carbon
 nanotubes are theoretically investigated using two different methods;
 continuous and lattice quantum theories.
 Both approaches show screening of a localized external perturbation
 with an effective screening length of the order of the nanotube
 diameter. Calculating the nonlinear deformation 
 of the local density of states near the
 charged perturbation, we show that the perturbative effects of the
 massive bands are effectively canceled by direct massive band
 interactions, such that a good agreement between the two methods can be
 achieved. 
 The effective screening is important in nanoscale integration of
 nanotube-based electronic devices.
\end{abstract}

\maketitle

\section{Introduction}

The response of the electrons in carbon nanotubes
(CNT's~\cite{Iijima,SDD}) to external charged perturbations is governed 
by the Coulomb interaction~\cite{Kane,EG,YO}.    
Several aspects of the Coulomb interaction are already seen in some
experiments of charging~\cite{Tans1,Tans2,Postma} and of temperature
dependent resistivity~\cite{Bockrath} which are consistent with current
understanding of low-energy excitations in metallic CNT's. 
The low-energy excitations in the metallic CNT's are theoretically
modeled by {\it massless} bands (linear dispersion bands) with the
Coulomb interaction~\cite{Kane,EG,YO}. 
However when we consider very localized perturbation, it is not clear if
the low energy theories describe the physics correctly. 
Because short distance physics corresponds to high energy, {\it massive}
bands (other sub-bands except the massless bands) would come into
dynamics~\cite{LC}.
From the application point of view, understanding the response of CNT's
to localized perturbations (e.g., individual dopant atoms) is crucial as 
it provides the extents of the depletion region in nanotube-based
electronic devices.
The importance of this issue is observed by considering
the fact that in any nano-electronics application of nanotubes, 
different functional parts of the device, e.g., drain, source, etc., 
should be doped differently. In order to achieve nanoscale 
integration, these differently doped parts should be 
separated from each other by distances of the order of 
a few nanometers. This would not be possible if the 
depletion-layer length is not of the order of a few nanometers itself. 
In this letter, taking massive band effects explicitly into account, we
focus on the screening of an external charge localized on the surface of
the metallic CNT's.

It is well-known that in lower dimensional materials in which the
internal electrons can move only in one or two dimensions, asymptotic
behavior of the potential produced by an external charge does not become
an exponential as a function of distance from the external charge.  
Examples of these lower dimensional materials are graphite sheet and
carbon nanotubes. 
DiVincenzo and Mele~\cite{DM} analyzed the induced charge around an
external charge on the graphite sheet and found that the external charge
attracts some induced charge with a characteristic length scale of the
system which they called ``effective screening length''. 
A similar definition is used to describe the effective screening in
metallic carbon nanotubes by Lin and Chuu~\cite{LC}. 
In the present paper, we study the details of the pattern of induced charge
around an external charge on the metallic nanotubes 
in order to examine the effects of massive bands on screening.

The high-energy excitations can in principle affect the screening
phenomena through two different mechanisms; adding massive-fermion loop 
corrections to the interactions of massless bands and direct interaction
of massive bands.
Two methods to deal with these mechanisms are investigated.
One is a quantum field theoretical method dealing with the massless bands
and the one-dimensional long-range Coulomb interaction~\cite{Sasaki},
with one-loop vacuum polarization correction by the massive bands.
The other is a lattice model employing a fully self-consistent
tight-binding approach including the interactions of massive
bands~\cite{Farajian}.
It is the purpose of the present work to compare the results of 
these two different methods and investigate the screening nature in
CNT's. 
Especially we shall compare the induced charge distribution around 
the external charge.
Our results are advantageous compared to the previous works on screening
in metallic nanotubes \cite{LC}, as we include an effective thickness for
the nanotube wall, and derive the dependence of our results on the
corresponding cut-off parameter. 
This inclusion is crucial in reaching an agreement between the
continuous and lattice results. 
Additionally, in our self-consistent tight-binding model there is 
no linear response assumption, and the deformation of local
density of states (LDOS), due to the creation of localized
states, are included. 
We also explicitly show how high-energy electron/hole excitations
respond differently to the presence of a charged perturbation.

\section{Continuous field theory: One-loop massive fermion corrections}

We begin with the quantum (continuous) field theoretical approach to the
screening effect. 
The $\pi$-electrons in CNT's are described by two-component fermion
fields $\psi_{Ns}$ specified by the band index $N (\in \{ 1 ,\cdots \})$
and spin $s (\in \{ \uparrow,\downarrow \})$. 
The Hamiltonian density of this system is given by ${\cal H} ={\cal H}_F
+ {\cal H}_C$ where ${\cal H}_F$ is the free (unperturbed) part and
${\cal H}_C$ is the perturbed part including the Coulomb interactions.
If we include only tight-binding interactions, then ${\cal H}_F =
\sum_{Ns} {\cal H}_N$ where 
\begin{equation}
 {\cal H}_N = \psi_{Ns}^\dagger V_\pi 
  \pmatrix{ 0 &  z(k^N_x,\hat{k}_y) \cr z^*(k^N_x,\hat{k}_y) & 0 } \psi_{Ns}.
\end{equation}
$V_\pi$ is the hopping integral, $\hat{k}_y = -i\partial_y$ is the wave
vector along the tubule axis, the wave vector around the circumference
$a k_x^N = - \frac{2\pi}{\sqrt{3}} + \frac{2a}{d}N$ is discretized by
the periodic boundary condition where $d$ is the diameter of the CNT and 
$a$ the nearest-neighbor C-C spacing.
Here we take the case of metallic zigzag CNT's in which the maximum
number of $N$ is $N_0$ (number of hexagons along the circumference);
armchair case is analyzed similarly. 
The function $z(k_x,k_y)$ is given by $z(k_x,k_y) = \sum_i e^{i k \cdot
u_i}$. 
The vector $u_i$ is a triad of vectors pointing in the direction of the
nearest neighbors of a carbon site~\cite{Sasaki}.  
The free Hamiltonian densities ${\cal H}_{N_0/3}$ and ${\cal
H}_{5N_0/3}$ describe massless fermions (electrons in massless bands)
and the others are for massive fermions in the metallic zigzag CNT.
The energy eigenvalues of ${\cal H}_N$ are given by $\pm V_\pi
|z(k_x^N,k_y)|$ which reproduce a standard band structure~\cite{SDD}. 
These fermions interact with each other via the long-range Coulomb
interaction: 
\begin{equation}
 {\cal H}_C = \frac{1}{2} \int_D  J(y)V(y-y')J(y') dy dy',
  \label{eq:coulomb}
\end{equation}
where $V(y)$ is the unscreened Coulomb potential that we will specify
shortly, and the internal charge density $J$ is given by the sum of
individual fermion densities: $J = \sum_{Ns} J_{Ns}$ where $J_{Ns}
\equiv \psi^\dagger_{Ns} \psi_{Ns}$.  
Here we denote the integral region $D = [0:L] $ and $L $ is the system
length.

Low-energy excitations in metallic CNT's are described by a field theory
of four massless fermions interacting with each other and with massive
fermions through the long-range Coulomb interaction. 
The massive fermions modify the Coulomb potential between massless fermions
because of the vacuum polarization. 
We estimate this effect using one-loop perturbation. 
The modified Fourier component of the Coulomb potential is given by
$\bar{\beta}_n = \frac{\beta_n}{1-\beta_n T^{\it ms}_n}$, where
$\beta_n$ is the $n$-th Fourier component of the unscreened Coulomb
potential, $V(x) = \sum_n \beta_n e^{-i2\pi n x/L}$, and $T^{\it
ms}_n (\le 0)$ in the denominator denotes the 
one-loop contribution of the {\it
massive} fermions~\cite{SFMK}. 
The unscreened Coulomb potential along the axis is~\cite{EG}
\begin{equation}
 V(x) = \frac{e^2}{4\pi \sqrt{|x|^2 + d^2 + a_z^2}}
  \frac{2}{\pi} K \left( \frac{d}{\sqrt{|x|^2 + d^2 + a_z^2}} \right),
  \label{eq:potential}
\end{equation}
with $K(z)$ being the complete elliptic integral of the first kind and 
$a_z  = 1.3[{\rm \AA}]$ a cutoff length which corresponds to the Hubbard
coefficient $U_H = 4 V_\pi $ in the lattice model~(Eq.\ref{eq:lattice3}).
The massive fermions are observed to weaken the Coulomb interaction
among massless fermions as is naturally expected. 
Because the Coulomb potential is modified by the vacuum polarization
effects, the massless fermions interact with each other via the screened
Coulomb potential whose Fourier components are given by $\bar{\beta}_n$.

\begin{figure}[htbp]
 \begin{center}
  \psfrag{x}{$[{\rm \AA}]$}
  \psfrag{y}{\hspace{-0.6cm}Total induced charge per ring $[e]$}
  \psfrag{a}{\hspace{-0.6cm} \scriptsize conti-armchair}
  \psfrag{b}{\hspace{-0.6cm} \scriptsize conti-armchair}
  \psfrag{B}{\hspace{-0.5cm} \scriptsize +massive loop}
  \psfrag{c}{\hspace{-0.5cm} \scriptsize lattice-armchair}
  \psfrag{D}{$D$}
  \psfrag{W}{$W$}
  \includegraphics[scale=0.6]{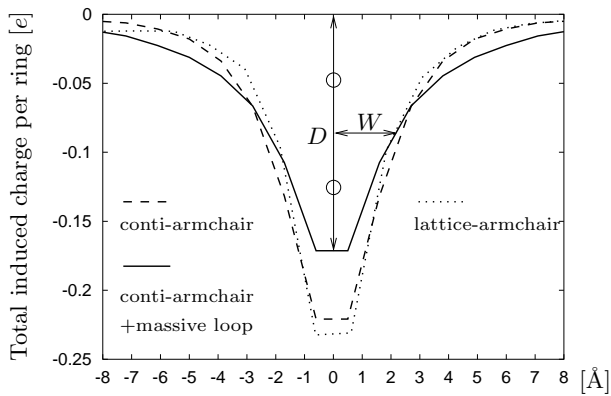}
 \end{center}
 \caption{Total induced charge (including spin degeneracy) around the
 point external charge for the (5,5) armchair tube. 
 The dotted line is obtained from the lattice model. 
 The solid and dashed lines are the continuous field theoretical
 results. 
 The dashed one is for the calculation using unscreened Coulomb
 potential $\beta_n$, and the solid one is for ${\bar \beta}_n$.
 The screening length is about the diameter of the nanotube. 
 Similar results are obtained for the zigzag (9,0) tube.}
 \label{fig:induced}
\end{figure}

An external charge can be included by replacing $J$ with $J+J^{ex}$ in
the Coulomb interactions (Eq.~\ref{eq:coulomb}), where $J^{ex}$ is a
c-number whose integral over the nanotube length is equal to the total
external charge. 
Here we take a point external charge distribution which is modeled by
$J^{ex}=\delta(x-x_0)$ where $x_0$ is the position of the external
charge. 
To calculate the induced charge density, we use one-loop approximation
which gives an exact result in the case of strictly linear dispersion  
relation~\cite{Sasaki}.
In Fig.~\ref{fig:induced}, the induced charge distributions around the
external point charge for both the unscreened potential and the screened
one are depicted. 
These are obtained using the following formula~\cite{SFMK,cut}:
\begin{equation}
 \langle J(x) \rangle =
  \sum_{n > 0}\frac{T^{\it ml}_n \bar{\beta}_n}{1-T^{\it ml}_n \bar{\beta_n}} 
  \frac{2}{L}
  \cos \left( \frac{2\pi n}{L}(x-x_0) \right),
  \label{eq:induced}
\end{equation}
where $T_n^{\it ml}$ is the one-loop amplitude of {\it massless} bands.
It should be noted that the effective screening length is about the
diameter of the nanotube for both cases and is not sensitive to the
system length~\cite{Sasaki}.

What we ignore in the above formula of induced charge density is 
(a) higher-order perturbative effects of massive loops on the Coulomb
interaction between massless fermions, and (b) induced charge due to
massive fermions (direct massive-band effect).
Lin and Chuu~\cite{LC} included the latter effect by adding 
$T^{\it ms}_n$ to $T^{\it ml}_n$ in the numerator of 
Eq.(\ref{eq:induced}). This assumption, however, can only result in a linear 
relationship between the external and induced charges, while
it is shown that linear-response theory is not sufficient
at least in the case of screening in graphite\cite{DM}.

\section{Self-consistent tight-binding: Nonlinear screening}

Up to now, only the massive-fermion corrections to the
{\it{massless}}-band interactions, through the leading terms of vacuum
polarization, were considered. 
We now turn to the {\it{massive}}-band effects on the screening
phenomena, and investigate the influence of direct massive-band
interactions. 
This is accomplished by using a lattice model to describe the nanotube.
The model is essentially a tight-binding one~\cite{Farajian} which is
generalized by including long-range Coulomb
interactions~\cite{Harigaya}.  

Within this model, the Hamiltonian of a perfect, infinite nanotube is
written as $H = H_F + H_C$ with  
\begin{eqnarray}
 &&
  H_F = \sum_{\langle i,j \rangle} 
  V_\pi \, c_i^\dagger c_j ,\\
 &&
  H_C = \sum_{i j}  
  \frac{\delta n_j}{\sqrt{r_{ij}^2 + U_H^{-2}}} 
  \, c_i^\dagger c_i ,
  \label{eq:lattice3}
\end{eqnarray}
where $U_H$ is the Hubbard parameter used in describing on-site
interactions.
Using the experimental estimation of $U_H$ for carbon~\cite{pearson},
we set $U_H = 4 V_\pi$. 
$U_H$ (corresponding to to the cut-off $a_z$)
is the only parameter of our model and we examine
the dependence of our results on this parameter, by considering
several cases of $U_H$ (or $a_z$).
In Eq.\ref{eq:lattice3}, $\delta n_i = n_i - n^0_i$ is the change in the
self-consistent and screened occupation number $n_i$ at site $i$, as
compared to the occupation imposed by external perturbation, $n^0_i$. 
Therefore, $\delta n_i$ would be the induced charge at site $i$, in
response to the external perturbation. 
We are concerned with localized perturbation, which is modeled by
modifying $n^0_i$ as $n^0_i \rightarrow n^0_i + n^{ex}_i$.

\begin{figure}[htbp]
 \begin{center}
  \psfrag{a}{$\frac{U_H}{V_\pi}\sim a_z^{-1}$}
  \psfrag{b}{\hspace{-0.0cm}Ratio $D/W$ [{\rm e/\AA}]}
  \psfrag{z}{\hspace{-0.3cm} \scriptsize conti-armchair}
  \psfrag{y}{\hspace{-0.3cm} \scriptsize conti-armchair+massive}
  \psfrag{x}{\hspace{-0.3cm} \scriptsize lattice-armchair}
  \includegraphics[scale=0.6]{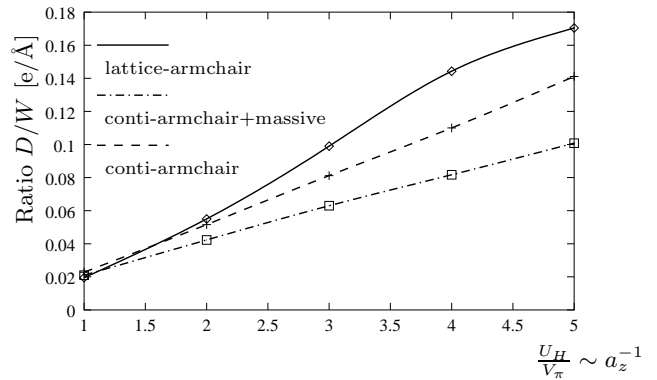}
 \end{center}
 \caption{The cutoff, $U_H$ (or $a_z$), dependence of the ratio defined by
 $D/W$, where $D$ is the peak of the induced charge at the position of
 the external charge and $W$ is the one-half of the width at the half
 peak (see Fig. \ref{fig:induced}).  
 The results show that bigger $U_H$ (smaller $a_z$) gives smaller
 screening length. 
 Experimentally, $U_H$ was reported \cite{pearson} to be between 2.4 and  
 4 $V_\pi$.}
 \label{fig:ratio}
\end{figure}

We distinguish three different parts in the infinite
nanotube: An unperturbed semi-infinite part to the left, a
perturbed finite part in the middle, and an
unperturbed semi-infinite part to the right. 
In the unperturbed parts, $n^0_i = 0.5$ 
(excluding the spin degeneracy) and
$\delta n_i = 0$. This implies that the length of the perturbed 
finite part,
for which $\delta n_i$'s are assumed to be nonzero, is
chosen to be long enough such that the external perturbation is fully
screened within this part of the nanotube.
For concreteness, we assume 
that the localized external
perturbation only modifies the sites 
of the two carbon rings of the nanotube 
by adding $n^{ex}_i$'s to the  corresponding $n^0_i$'s. (A ring
consists of all the carbon atoms with the same longitudinal
coordinate along the nanotube axis.)
The two rings with nonzero $n^{ex}_i$'s are assumed to be 
at the middle of the perturbed finite part 
of the nanotube, which, in turn, is taken to consist of 16 carbon rings.
This number of carbon rings has proven ~\cite{Farajian}
to be long enough in order to derive the converged induced
charges, as increasing the number of rings has had no effect
on the self-consistent occupations. 

The self-consistent procedure of calculating the induced 
charges $\delta n_i$'s proceeds as follows ~\cite{Farajian}:
Starting with an initial guess of zero $\delta n_i$'s in the
Hamiltonian Eq.\ref{eq:lattice3}, the 
surface Green's functions, $G_L$ and $G_R$, of the two semi-infinite
unperturbed parts to the left and right of the perturbed region,  
are calculated\cite{lopez-sancho}. These surface Green's
functions are then attached to the Green's function of
the perturbed middle part, $G_P = (z - H_P)^{-1}$, where
$z$ is the complex energy and $H_P$ is the Hamiltonian of the 
perturbed region, in order to 
obtain the total Green's function of the system, $G_T$, projected 
onto the perturbed region\cite{lee,mckinnon,munoz,garcia,datta,chico}:
\begin{equation}
G_T=(G_P^{-1} - V_{PL} G_L V_{LP} - V_{PR} G_R V_{RP})^{-1} .
  \label{eq:lattice4}
\end{equation}
  
\noindent Here, $V$ matrices indicate the coupling between
the perturbed region and the first unit cells of the unperturbed
parts to the left and right.
Using Eq.\ref{eq:lattice4}, 
the LDOS at site $i$ in the perturbed region,
$g_i(E) = [G_T(i,i) - G_T^*(i,i)]/2\pi$, is derived.  
The induced charges can be obtained through
integrating $g_i(E)$'s. $\delta n_i$'s are then used
to initialize the Hamiltonian for another round of
$\delta n_i$'s calculation. This procedure is continued
until the induced charges are determined self-consistency.  
Throughout the self-consistent calculation, the chemical potential
of the system is assumed to be that of the unperturbed nanotube,
as after screening the whole system would be in thermal 
equilibrium and would have a unique chemical potential. Furthermore,
the external perturbation being localized and finite, 
the unique chemical potential would be that of the 
unperturbed semi-infinite parts of the nanotube, which
act like reservoirs attached to the perturbed finite part.

As specific examples, an armchair (5,5) and a zigzag (9,0)
nanotube are considered, whose diameters are 6.78 and 7.04 [{\AA}]. 
The external perturbation of the middle two rings
of the perturbed part of these tubes are defined as 
$n^{ex}_i = 0.0250$ and $n^{ex}_i = 0.0278$, for the armchair
and zigzag tubes, respectively. 
This external perturbation amounts to adding a total charge of 
$1.0 e$ to the middle two rings
for both the armchair and the zigzag tubes. 

Comparing the self-consistent procedure used in the tight-binding 
model of the present work with the tight-binding approach of
Lin and Chuu \cite{LC}, we notice that in our model there
is no restriction on the shape of the LDOS within the perturbed
region. In fact, the very shape of the LDOS is calculated
self-consistently, as described above. Lin and Chuu's approach,
however, assumes fixed energy dispersion relations for the 
whole nanotube, and restricts the modifications due to the 
charged perturbation to the linear response theory via dielectric
constant and response function. Our model is therefore advantageous,
as it properly takes into account the nonlinear effects of the 
localized states which are produced within the perturbed 
region. This is explicitly observed from the LDOS curves
in the perturbed region, as will be shown shortly.

\section{Comparison and discussion}

The resultant induced charges for the (5,5) 
armchair tube are depicted in 
Fig.~\ref{fig:induced}. It is observed
that the effective screening length is of the order of the 
nanotube diameter, in all the different calculations.
The length also depends on the cutoff $a_z$ or 
the Hubbard coefficient $U_H$. The cutoff dependence of induced charge
pattern is plotted for several $U_H$ values in Fig.~\ref{fig:ratio}.
The main features of the 
induced charge pattern are observed to hold for the
whole range of experimentally reported $U_H$~\cite{pearson}. 
From Fig.~\ref{fig:induced} we further see that the induced charges 
obtained by including only the massless bands 
in the continuous field-theoretical calculation are in 
qualitative agreement with the results of 
lattice calculation, which, in addition to the massless bands, 
includes the effects
of direct massive bands interactions. Interestingly,
including the massive-loop corrections to massless bands
interactions does not result in better agreement
with the results of lattice model. This suggests that,
within the continuous model, the massive-loop corrections to
the massless band interactions somehow cancel the effects of
direct massive band interactions, such that the results 
of a calculation based only on massless bands turn out to
be in agreement with the results of the lattice model. 
In other words, in the continuous approach, 
the induced charge density is given by the sum of two densities as
$\langle J \rangle = \langle J_{\rm massless} \rangle 
        + \langle J_{\rm massive} \rangle$ 
where $\langle J_{\rm massless} \rangle$ is defined 
in Eq.\ref{eq:induced}. $\langle J_{\rm massive} \rangle$ is 
an induced charge density due to massive bands themselves, and 
is what we have ignored. Perturbative inclusion of massive 
bands reduces the peak of the former charge
density and results in a more appreciable difference with the lattice results.
Considering overall agreement of the lattice results and continuous field 
theoretical results based on the massless bands, it is natural to expect 
that $\langle J_{\rm massive} \rangle$ partially cancels 
the perturbative effects of the
massive bands on the massless bands interaction,
such that the net induced charge density 
is basically decided by the massless bands.

In addition to the calculations for the 
armchair tube, we have performed similar calculations 
on the (9,0) zigzag tube. The results of the continuous
model~\cite{Sasaki} show the same features, as those of the 
armchair one, when compared to the results of lattice model. 
This indicates that the agreement between the 
continuous massless results and lattice results is 
not accidental.   
It is remarkable that in both the continuous and lattice 
calculations the peak value of the induced charge shows
a reduction of $\sim$ 15\%, when one compares zigzag results
with armchair results.

\begin{figure}[ttbp]
\begin{center}
\psfrag{a}{$[\frac{E}{V_\pi}]$}
\psfrag{b}{\hspace{-0.7cm} Self-consistent electron LDOS}
\includegraphics[scale=0.6]{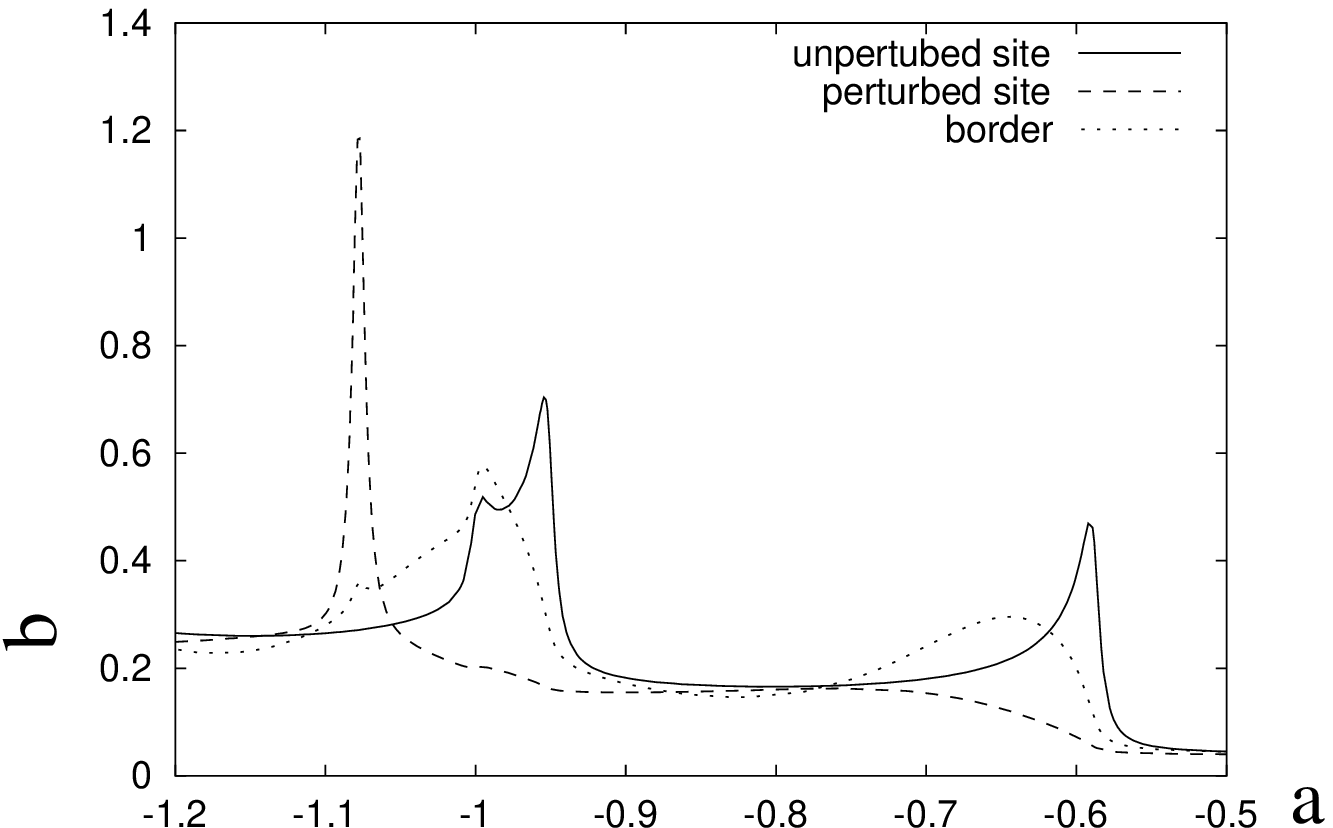}
\psfrag{b}{\hspace{-0.4cm} Self-consistent hole LDOS}
\includegraphics[scale=0.6]{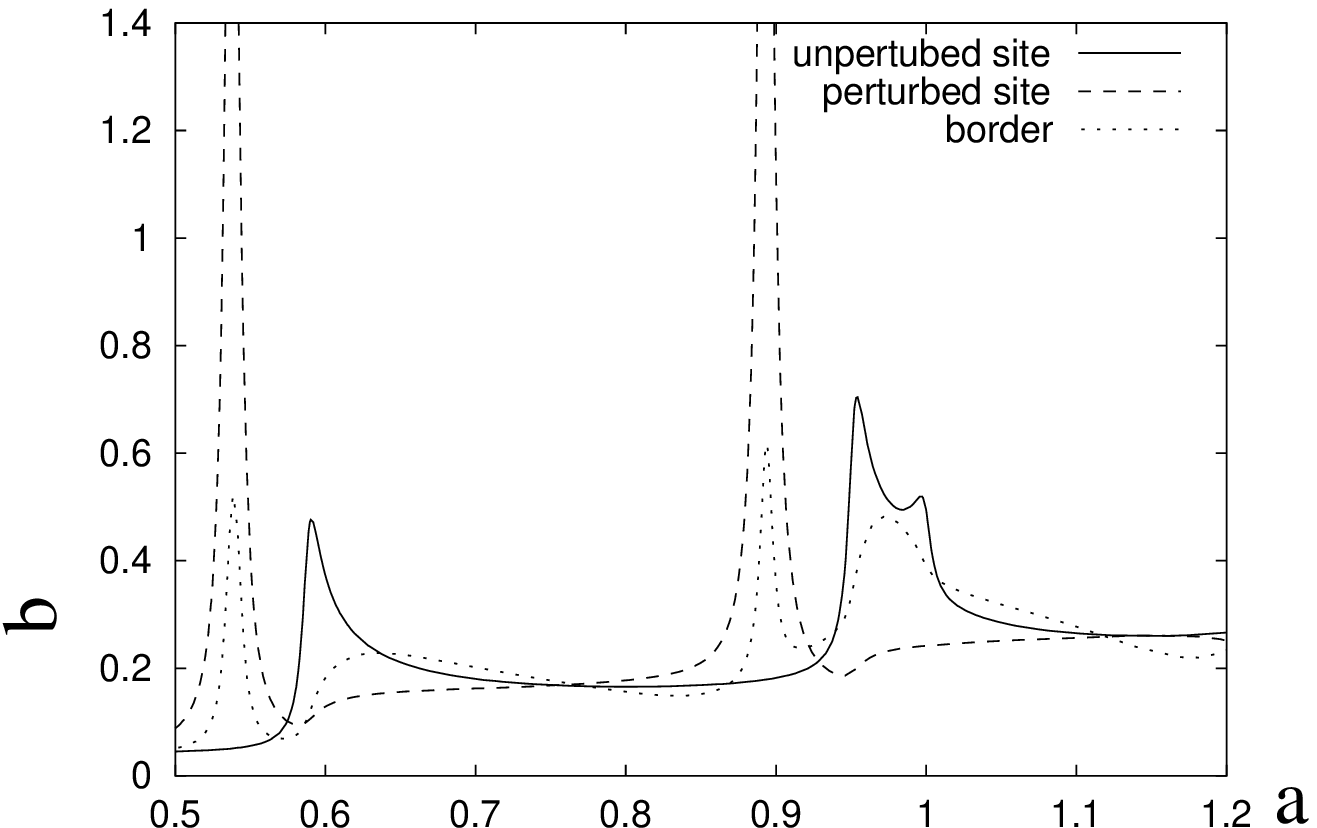}
\end{center}
\caption{Self-consistent electron (top) and 
hole (bottom) LDOS. Fermi energy is located at zero.}
\label{fig:ldos}
\end{figure}

In order to see the LDOS deformation, the creation of localized
states within the perturbed region, and the effects 
of massive bands more clearly, 
we show the self-consistent 
LDOS's for a typical unperturbed
site (solid line), a site located at 
one of the rings perturbed externally (dashed line), 
and a site, of the perturbed
portion of the nanotube, located at the furthest ring 
relative to the middle two rings (dotted line) in Fig.~\ref{fig:ldos}. 
The latter site indicates the border between the perturbed
finite part of the tube and one of the unperturbed semi-infinite
parts. In Fig.~\ref{fig:ldos}, enlarged typical
regions of the electron (hole) LDOS's are depicted.
It is clearly seen that the densities of 
massive states in the perturbed region are
different from the corresponding densities of the unperturbed
region. However, as one expects, the difference is much more 
pronounced near the location of external perturbation than
far from it. This is indeed an indication of localized-states 
creation. The border LDOS shows how the 
extremely perturbed LDOS near the external perturbation
transforms rather smoothly, as one moves toward the
border of the perturbed region, into the unperturbed LDOS.
Another very interesting feature is the difference in the
behavior of the electron states as compared to the hole states.
For the external perturbation mentioned above, namely
{\it{addition}} of one extra electron to the middle two rings,
it is seen that the hole states tend to accumulate at lower
energies, while the electron states are pushed toward higher
energies. This indicates the decisive effect of the Coulomb
interactions among the massive fermion- (both electron- and hole- ) 
states. One should notice that, due to the rotational symmetry
selection rule ~\cite{Farajian}, the electron/hole states 
belonging to a certain band
are not allowed to jump over to another band with higher/lower
energy. But {\it{within each band}} the electron states
tend to concentrate at higher energies, and the hole states
tend to concentrate at lower energies. This is indeed the
reason for the enhanced/decreased densities at band edges. 
    
\section{Conclusions}

In summary, we have analyzed the charge screening in metallic nanotubes
using two different methods, including explicitly the massive bands
effects, based on lattice and continuous models.
In both methods, a localized external charge is shown to be screened
with an effective screening length about the diameter of the tube.
The local density of states, calculated by the lattice model, indicates
the formation of the localized states within the perturbed region,
and shows that the massive bands are 
strongly affected by the external charge. 
As for the continuous field theoretical model, we have included the
massive bands one-loop corrections in massless bands interactions.
Although non-perturbative effects of massive bands were not considered,
the results show basically the same pattern as that of the lattice model
results.  
This shows that the perturbative corrections of massive bands are
efficiently canceled by the induced charge due to massive bands
themselves.
The short effective screening length reported here makes it possible, at
least theoretically, to achieve nanoscale integration of nanotube-based
electronic devices. \\ 

\begin{acknowledgments}
A.A.F., H.M. and Y.K. are supported by the Special Coordination Funds 
of the Ministry of Education, Culture, Sports, Science and Technology
of the Japanese government.
\end{acknowledgments}



\begin{thebibliography}{99}
\bibitem{cut}
	We set $n$ equal to $L/\sqrt{3}a$ for the armchair CNT. 
	This cutoff may cause an oscillatory pattern~\cite{Sasaki} of
	the induced charge density for the  Coulomb interaction with
	shorter cutoff length $a_z$.
\end{thebibliography}
\end{document}